\documentclass[twocolumn,showpacs,preprintnumbers,amsmath,amssymb]{revtex4}
\usepackage{epsfig}
\usepackage{dcolumn}
\usepackage{bm}

\newcommand{\remove}[1]{}
\newcommand{\dd}{\mathrm{d}}

\def\be{\begin{equation}}
\def\ee{\end{equation}}

\newcommand{\beq}{\begin{equation}}
\newcommand{\eeq}{\end{equation}}
\newcommand{\beqa}{\begin{eqnarray}}
\newcommand{\eeqa}{\end{eqnarray}}

\renewcommand{\pl}{\partial}
\newcommand{\lag}{\langle}
\newcommand{\rag}{\rangle}

\newcommand{\ii}{{\rm i}}

\newcommand{\vn}{{\bf n}}

\newcommand{\vv}{{\bf v}}

\newcommand{\vx}{{\bf x}}
\newcommand{\vk}{{\bf k}}

\newcommand{\vq}{{\bf q}}

\newcommand{\vF}{{\bf F}}
\newcommand{\vPsi}{{\bf \Psi}}

\newcommand{\tdelta}{{\tilde{\delta}}}

\newcommand{\tPhi}{{\tilde{\Phi}}}
\newcommand{\tpsi}{{\tilde{\psi}}}

\newcommand{\trho}{{\tilde{\rho}}}

\newcommand{\cD}{{\cal D}}

\newcommand{\cG}{{\cal G}}
\newcommand{\cH}{{\cal H}}

\newcommand{\cO}{{\cal O}}

\newcommand{\rhob}{\overline{\rho}}
\newcommand{\Om}{\Omega_{\rm m}}

\newcommand{\bea}{\begin{array}}
\newcommand{\ea}{\end{array}}

\newcommand{\MPl}{M_{\rm Pl}}

\begin{document}

\title{Kinematic consistency relations of large-scale structures}

\author{Patrick Valageas}
\affiliation{Institut de Physique Th\'eorique,\\
CEA, IPhT, F-91191 Gif-sur-Yvette, C\'edex, France\\
CNRS, URA 2306, F-91191 Gif-sur-Yvette, C\'edex, France}
\vspace{.2 cm}

\date{\today}
\vspace{.2 cm}

\begin{abstract}
We describe how the kinematic consistency relations satisfied by density correlations
of the large-scale structures of the Universe can be derived within the usual
Newtonian framework. These relations express a kinematic effect and show 
how the $(\ell+n)$-density correlation factors in terms of the $n$-point correlation
and $\ell$ linear power spectrum factors, in the limit where the $\ell$ soft wave
numbers become linear and much smaller than the $n$ other wave
numbers. We describe how these relations extend to multifluid cases.
In the standard cosmology, these consistency relations derive from the
equivalence principle. A detection of their violation would indicate non-Gaussian 
initial conditions, non-negligible decaying modes, or
a modification of gravity that does not converge to General Relativity on large
scales.

\end{abstract}

 \vskip2pc

\maketitle

\section{Introduction}
\label{Introduction}

The large-scale structure of the Universe is the main probe of the recent evolution
of the Universe and of the properties of still mysterious components such as
dark matter and dark energy. Unfortunately, even without considering the very
complex processes of galaxy and star formation and focusing on the large-scale
properties where gravity is the dominant driver, theoretical progress is difficult.
Large scales can be described by standard perturbative approaches
\cite{Goroff1986,Bernardeau2002},
which can be improved to some degree by using resummation schemes
\cite{Crocce2006a,Valageas2007,Pietroni2008,Bernardeau2008,Taruya2012,Pietroni2012,Crocce2012,Bernardeau2013,Valageas2013}.
However, these methods cannot reach the truly nonlinear regime where
shell-crossing effects become important 
\cite{Pueblas2009,Valageas2011a,Valageas2013a}.
Small scales are studied through numerical
simulations or phenomenological models \cite{Cooray2002}
that rely on information gained through these simulations.
However, these scales are very difficult to model with a high accuracy, even with 
simulations, because of the complexities of galaxy formation processes
and feedback effects such as AGN and supernovae outflows
\cite{Scannapieco2012,Bryan2013,Semboloni2013,Martizzi2013}.
Therefore, exact results that do not depend on the small-scale nonlinear physics
are very important.

Such results have been recently obtained in \cite{Kehagias2013,Peloso2013}
in the nonrelativistic limit, then extended in \cite{Creminelli2013} to the
relativistic case, and further explored in  
\cite{Peloso2013a,Creminelli2013a,Kehagias2013a}. 
These ``consistency relations'' express correlations between large-scale
linear modes and small-scale (even nonlinear) modes as a product of
the linear modes' power spectra with the small-scale correlation (at lowest
order over the scale ratio).
The great advantage of these relations is that they remain valid independently
of the small-scale physics, which can be highly nonlinear and involve astrophysical
processes such as star formation and supernovae outflows.
As nicely described in \cite{Creminelli2013}, within the context of General Relativity
and for standard scenarios, these consistency relations follow from
the equivalence principle. This ensures that small-scale structures are transported
without distortions by large-scale fluctuations, which at leading order correspond
to a constant gravitational force over the extent of the small-scale region.
Thus, these consistency relations express a kinematic effect and
describe how small scales are transported with time by large-scale gravitational
forces.

In this paper we present a simple and more explicit derivation (without using the
single-stream approximation) in the nonrelativistic
framework that is most often used for studies of large-scale structures. 
This also provides a generalization to an arbitrary number of soft wave numbers and
fluid components.
This allows us to distinguish which ingredients
are required for their validity. In particular, we recover the fact that the equivalence
principle is sufficient to guarantee the consistency relations, once we assume
Gaussian initial conditions and negligible decaying modes 
(more generally, a weaker form of scale separation is sufficient, but this extension
is not needed for realistic scenarios).

This paper is organized as follows.
We first derive the consistency relations in Sec.~\ref{consist}, within a very
general framework based on Gaussian initial conditions, using an assumption of
scale separation (which states that large-scale fluctuations have an almost uniform
impact on small-scale structures). 
We also consider the cases of an arbitrary number 
of soft wave numbers and of several fluid components.
Next, we discuss in Sec.~\ref{examples} the conditions of validity of these
consistency relations and we conclude in Sec.~\ref{Conclusion}.

\section{Consistency relations for density field correlations}
\label{consist}

\subsection{Correlation and response functions}
\label{corr-resp}

Let us consider a system fully determined by a field $\varphi(x)$,
which may be for instance the initial condition of a dynamical system
[in our case $\varphi$ will be the Fourier-space linear density contrast 
$\tdelta_{L0}(\vk)$ today]. We also consider quantities
$\{\rho_1,..,\rho_n\}$ that are functionals of the field $\varphi$
[in our case $\rho_i$ will be the Fourier-space nonlinear density contrast
$\tdelta(\vk_i,t_i)$ at wave number $\vk_i$ and time $t_i$].
Then, general relations between correlation functions and response functions
can be obtained from integrations by parts
\cite{Valageas2007a,Bernardeau2012}.
Thus, considering the Gaussian case where the statistical properties
of the field $\varphi(x)$ are defined by its two-point correlation
$C_0(x_1,x_2) =  \lag \varphi(x_1) \varphi(x_2)\rag$,
the mixed correlations can be written as the Gaussian average
\beqa
C^{\ell,n} & = & \lag \varphi(x_1) \dots \varphi(x_{\ell}) \,
\rho_1 \dots \rho_n \rag \nonumber \\
& = & \int\! \cD \varphi \; e^{ - (1/2) \varphi \cdot C_0^{-1} \cdot \varphi } \;
\varphi(x_1) \dots \varphi(x_{\ell}) \nonumber \\
&& \times \; \rho_1 \dots \rho_n .
\label{xin-1}
\eeqa
If the inverse correlation matrix satisfies $C_0^{-1}(x_i,x_j)=0$ for $i\neq j$,
we also have the functional derivatives
\beqa
\frac{\cD^{\ell} [ e^{-(1/2) \varphi \cdot C_0^{-1} \cdot \varphi } ]}
{\cD\varphi(x_1) .. \cD\varphi(x_{\ell})} 
& = & (-1)^{\ell} \, C_0^{-1}(x_1,x_1') \cdot \varphi(x_1') \nonumber \\
&& \hspace{-2.9cm} \times  \dots \times C_0^{-1}(x_{\ell},x_{\ell}') \cdot 
\varphi(x_{\ell}') \; e^{-(1/2) \varphi \cdot C_0^{-1} \cdot \varphi } .
\label{deriv1}
\eeqa
Therefore, we can write Eq.(\ref{xin-1}) as
\beqa
C^{\ell,n} & = & (-1)^{\ell} C_0(x_1,x_1') .. C_0(x_{\ell},x_{\ell}') \cdot
\int\! \cD \varphi \; \rho_1 \dots \rho_n \nonumber \\
&& \times \frac{\cD^{\ell} [ e^{-(1/2) \varphi \cdot C_0^{-1} \cdot \varphi } ]}
{\cD\varphi(x_1') .. \cD\varphi(x_{\ell}')} \nonumber \\
& = & C_0(x_1,x_1') .. C_0(x_{\ell},x_{\ell}') \cdot \int\! \cD \varphi \; 
e^{ - (1/2) \varphi \cdot C_0^{-1} \cdot \varphi } \nonumber \\
&& \times 
\frac{\cD^{\ell} [ \rho_1 \dots \rho_n ]}
{\cD\varphi(x_1') .. \cD\varphi(x_{\ell}')} ,
\label{xin-2}
\eeqa
where we made $\ell$ integrations by parts. This gives the relation
\beq
C^{\ell,n}(x_1,..,x_{\ell}) = C_0(x_1,x_1') .. C_0(x_{\ell},x_{\ell}') \cdot
R^{\ell,n}(x_1',..,x_{\ell}')
\label{C-R-1}
\eeq
between the correlation $C^{\ell,n}$ and the response function $R^{\ell,n}$ defined
by
\beq
R^{\ell,n}(x_1,..,x_{\ell}) = \lag \frac{\cD^{\ell} [ \rho_1 \dots \rho_n ]}
{\cD\varphi(x_1) .. \cD\varphi(x_{\ell})} \rag .
\label{R-def}
\eeq

In the cosmological context, working in Fourier space, we take $\varphi$ as the
linear matter  density contrast today, $\tdelta_{L0}(\vk')$, and $\rho_i$ as the
nonlinear density contrast $\tdelta(\vk_i,t_i)$ at wave number $\vk_i$ and time
$t_i$, where $\delta=(\rho-\rhob)/\rhob$.
[The system is fully defined by $\tdelta_{L0}$ because we assume that the linear
decaying mode has had time to become negligible, so that $\varphi$ also
specifies the initial condition $\tdelta_{LI}=D_+(t_I) \tdelta_{L0}$
at the initial time $t_I \rightarrow 0$, where $D_+(t)$ is the linear growth rate.]
Then, the linear correlation function is
\beq
C_{L0}(\vk_1,\vk_2) = \lag \tdelta_{L0}(\vk_1) \tdelta_{L0}(\vk_2) \rag = 
P_{L0}(k_1) \delta_D(\vk_1+\vk_2) ,
\label{C0-def}
\eeq
where $P_{L0}$ is the linear matter power spectrum, with the inverse
\beq
C_{L0}^{-1}(\vk_1,\vk_2) = P_{L0}(k_1)^{-1} \delta_D(\vk_1+\vk_2) .
\label{C0m-def}
\eeq
Thus, if the wave numbers $\{\vk'_i\}$ satisfy $\vk'_i+\vk'_j\neq 0$ for all pairs
$\{i,j\}$, Eq.(\ref{C-R-1}) can be written as
\beqa
C^{\ell,n}(\vk'_1,..,\vk'_{\ell};\vk_1,t_1,..,\vk_n,t_n) & = & P_{L0}(k_1') .. P_{L0}(k_{\ell}')
\nonumber \\
&& \hspace{-3cm} \times R^{\ell,n}(-\vk'_1,..,-\vk'_{\ell};\vk_1,t_1,..,\vk_n,t_n) ,
\label{tC-tR-1}
\eeqa
where
\beq
C^{\ell,n}(\vk'_j;\vk_i,t_i) = \lag \tdelta_{L0}(\vk'_1) .. \tdelta_{L0}(\vk'_{\ell}) \, 
\tdelta(\vk_1,t_1) .. \tdelta(\vk_n,t_n) \rag
\label{tC-def}
\eeq
and
\beq
R^{\ell,n}(\vk'_j;\vk_i,t_i)  = \lag \frac{\cD^{\ell} [\tdelta(\vk_1,t_1) .. \tdelta(\vk_n,t_n)]}
{\cD\tdelta_{L0}(\vk'_1) .. \cD\tdelta_{L0}(\vk'_{\ell})} \rag .
\label{tR-def}
\eeq
In this paper, we denote all wave numbers associated with the initial field
$\tdelta_{L0}$ or soft wave numbers with a prime.

\subsection{Consistency relations}
\label{cons-rel}

Turning to a Lagrangian point of view, matter particles follow trajectories $\vx(\vq,t)$
labeled by their initial (Lagrangian) coordinate $\vq$. The conservation of matter
means that $(1+\delta) \dd\vx= \dd\vq$, and the Fourier-space density contrast
can also be written as
\beq
\tdelta(\vk,t) = \int \frac{\dd\vx}{(2\pi)^3} \; e^{-\ii\vk\cdot\vx} \delta(\vx,t) 
=  \int \frac{\dd\vq}{(2\pi)^3} \; e^{-\ii\vk\cdot\vx(\vq,t)} ,
\label{delta-q}
\eeq
where we discarded a Dirac factor that does not contribute for $\vk \neq 0$.
Therefore, the density-contrast response functions write as
\beqa
R^{\ell,n} & = & \lag  \int \frac{\dd\vq_1 ..\dd\vq_n}{(2\pi)^{3n}} \; 
\frac{\cD^{\ell}}{\cD\tdelta_{L0}(\vk'_1) .. \cD\tdelta_{L0}(\vk'_{\ell})} \nonumber \\
&& \times \; e^{-\ii \vk_1 \cdot (\vq_1+\vPsi_1) - .. - \ii \vk_n \cdot (\vq_n+\vPsi_n)} \rag ,
\label{R-psi-1}
\eeqa
where we introduced the displacement field $\vPsi(\vq,t) = \vx(\vq,t)-\vq$.

Let us first consider the case $\ell=1$, where Eq.(\ref{R-psi-1}) reads
\beqa
R^{1,n} & = & - \ii \, \lag  \int \frac{\dd\vq_1 ..\dd\vq_n}{(2\pi)^{3n}} \; 
\sum_{i=1}^n \vk_i\cdot \frac{\cD\vPsi_i}{\cD\tdelta_{L0}(\vk')} \nonumber \\
&& \times \; e^{-\ii \vk_1 \cdot (\vq_1+\vPsi_1) - .. - \ii \vk_n \cdot (\vq_n+\vPsi_n)} \rag .
\label{R-psi-1-1}
\eeqa
We now assume that, if we look at a fixed region of size $L$ and volume $V=L^3$,
a perturbation to the initial conditions $\tdelta_{L0}$ at a larger-scale linear wave
number $k' \ll 1/L$ gives rise to an almost uniform displacement of the small-size
region, at leading order over $(k'L)$. Thus, we write
\beq
k'\rightarrow 0, \;\;\; k' L \ll 1 : \;\;\; \frac{\cD\vPsi(\vq)}{\cD\tdelta_{L0}(\vk')} \simeq
\int_V \frac{\dd\vq'}{V} \; \frac{\cD\vPsi(\vq')}{\cD\tdelta_{L0}(\vk')} ,
\label{dpsi-delta-1}
\eeq
where we integrate over a volume $V$ centered on $\vq$.
Next, in the limit $k'\rightarrow 0$ we can take for instance $L \sim 1/\sqrt{k'}$
so that the size $L$ also goes to infinity (while keeping it much smaller than $1/k'$).
Then, we also assume that on large scales we recover the linear theory,
\beq
k \rightarrow 0: \;\;\;  \tilde{\vPsi}(\vk) \rightarrow \tilde{\vPsi}_L(\vk) ,
\label{psi-linear-1}
\eeq
so that Eq.(\ref{dpsi-delta-1}) implies
\beq
k'\rightarrow 0 : \;\;\; \frac{\cD\vPsi(\vq)}{\cD\tdelta_{L0}(\vk')} \rightarrow
\frac{\cD\vPsi_L(\vq)}{\cD\tdelta_{L0}(\vk')} .
\label{psi-psiL-1}
\eeq
On the other hand, the conservation of matter can also be expressed through the
continuity equation,
\beq
\frac{\pl \delta}{\pl\tau} + \nabla \cdot [ (1+\delta) \vv ] = 0 ,
\label{continuity}
\eeq
where $\tau=\int\dd t /a$ is the conformal time and $\vv$ the peculiar velocity
($\vv=\dd\vx/\dd\tau=\dd\vPsi/\dd\tau$). At linear order this gives 
$\pl_{\tau}\delta_L + \nabla\cdot\vv_L=0$, whence
\beq
\tilde{\vPsi}_L(\vk,\tau) = \ii \frac{\vk}{k^2} \, \tdelta_L(\vk,\tau) = 
\ii \frac{\vk}{k^2} \, D_+(k,\tau) \, \tdelta_{L0}(\vk) .
\label{v-delta-1}
\eeq
The linear growth rate of the density contrast $D_+(\tau)$ (which we normalize
to unity today) does not depend on scale
in the standard $\Lambda$-CDM cosmology, but this is no longer true in some
modified-gravity scenarios. Therefore, we include a possible $k$-dependence for
completeness.
Substituting into Eq.(\ref{psi-psiL-1}) we obtain
\beq
k' \rightarrow 0 : \;\;\; \frac{\cD\vPsi(\vq)}{\cD\tdelta_{L0}(\vk')} \rightarrow
\ii \frac{\vk'}{k'^2} \, \bar{D}_+(\tau) ,
\label{dpsi-k0-1}
\eeq
where we note with the overbar the low-$k$ limit of the linear growth rate,
$\bar{D}_+(\tau)= D_+(0,\tau)$. 
We postpone to Sec.~\ref{examples} a more explicit derivation of Eq.(\ref{dpsi-k0-1})
than the intuitive argument (\ref{dpsi-delta-1}), as well as the discussion of its validity, 
because we first wish to show how consistency relations
for arbitrary numbers of soft wave numbers and fluid components follow from this
property.

Then, using the expression (\ref{dpsi-k0-1}) in Eq.(\ref{R-psi-1-1}), we obtain
\beqa
R^{1,n}_{k'\rightarrow 0} & = & \lag  \int \frac{\dd\vq_1 ..\dd\vq_n}{(2\pi)^{3n}} \; 
\sum_{i=1}^n \frac{\vk_i\cdot\vk'}{k'^2} \bar{D}_+(t_i) \nonumber \\
&& \times \; e^{-\ii \vk_1 \cdot (\vq_1+\vPsi_1) - .. - \ii \vk_n \cdot (\vq_n+\vPsi_n)} \rag .
\label{R-1n-1}
\eeqa
Thus, the prefactor generated by the functional derivative in Eq.(\ref{R-psi-1-1}) has
a deterministic large-scale limit, which does not depend on the initial conditions,
and the statistical average gives [compare with Eq.(\ref{delta-q})]
\beq
R^{1,n}_{k'\rightarrow 0} = \lag \tdelta(\vk_1,t_1) .. \tdelta(\vk_n,t_n) \rag
\sum_{i=1}^n \frac{\vk_i\cdot\vk'}{k'^2} \bar{D}_+(t_i) .
\label{R-1n-2}
\eeq
Substituting into Eq.(\ref{tC-tR-1}) we obtain
\beqa
\lag \tdelta_{L0}(\vk') \, \tdelta(\vk_1,t_1) .. \tdelta(\vk_n,t_n) \rag'_{k'\rightarrow 0} \!\!
& = & \!\! - \! \sum_{i=1}^n \frac{\vk_i \! \cdot \! \vk'}{k'^2} \bar{D}_+(t_i) \;\;\; \nonumber \\
&& \hspace{-2.5cm} \times \, P_{L0}(k') \, \lag \tdelta(\vk_1,t_1) .. \tdelta(\vk_n,t_n) \rag' ,
\label{d0-dn-1}
\eeqa
Here and in the following, the prime in $\lag .. \rag'$ denotes that we removed the 
Dirac factor $\delta_D(\sum \vk_i)$ from the correlation functions.

The result (\ref{d0-dn-1}) can be extended at once to $\ell \geq 2$.
Indeed, each derivative $\cD/\cD\tdelta_{L0}(\vk'_i)$ in Eq.(\ref{R-psi-1})
generates a constant prefactor, given by Eq.(\ref{dpsi-k0-1}), which is not affected by
the next derivatives. This yields
\beq
R^{\ell,n}_{k_j'\rightarrow 0} = \lag \tdelta(\vk_1,t_1) .. \tdelta(\vk_n,t_n) \rag
\prod_{j=1}^{\ell}  \left( \sum_{i=1}^n \frac{\vk_i\cdot\vk_j'}{k_j'^2} 
\bar{D}_+(t_i) \right) .
\label{R-ln-1}
\eeq
Substituting into Eq.(\ref{tC-tR-1}) we obtain
\beqa
\lag \tdelta_{L0}(\vk'_1) .. \tdelta_{L0}(\vk'_{\ell}) \, \tdelta(\vk_1,t_1) .. \tdelta(\vk_n,t_n) 
\rag'_{k'_j\rightarrow 0} & = & \nonumber \\
&& \hspace{-5cm} \prod_{j=1}^{\ell}  \left( - P_{L0}(k_j') \sum_{i=1}^n 
\frac{\vk_i\cdot\vk_j'}{k_j'^2} \bar{D}_+(t_i) \right) \nonumber \\
&& \hspace{-5cm} \times \; \lag \tdelta(\vk_1,t_1) .. \tdelta(\vk_n,t_n) \rag' ,
\label{dl-dn-1}
\eeqa
where the soft wave numbers must satisfy the condition $\vk_i'+\vk_j' \neq 0$ for
all pairs $\{i,j\}$.
Since on large scales we have 
$\tdelta_L(\vk,t) \simeq D_+(k,t) \tdelta_{L0}(\vk)$, Eq.(\ref{dl-dn-1}) also
writes as
\beqa
\lag \tdelta(\vk_1',t_1') .. \tdelta(\vk_{\ell}',t_{\ell}') \, \tdelta(\vk_1,t_1) .. \tdelta(\vk_n,t_n) 
\rag'_{k_j'\rightarrow 0} & = & \nonumber \\
&& \hspace{-6cm} P_L(k_1',t_1') .. P_L(k_{\ell}',t_{\ell}') \,
\lag \tdelta(\vk_1,t_1) .. \tdelta(\vk_n,t_n) \rag'  \nonumber \\
&& \hspace{-6cm} \times \prod_{j=1}^{\ell}  \left( - \sum_{i=1}^n 
\frac{\vk_i\cdot\vk_j'}{k_j'^{2}} \frac{\bar{D}_+(t_i)}{\bar{D}_+(t_j')} \right) ,
\label{dl-dn-2}
\eeqa
with the condition $\vk_i'+\vk_j'\neq 0$ for $i\neq j$.
Thus, Eq.(\ref{dl-dn-2}) shows how the density correlation functions 
$\lag \tdelta_1 .. \tdelta_{\ell+n}\rag$ factorize when $\ell$ wave numbers 
are within the linear regime and become very small as compared with the fixed
$n$ other wave numbers.
This generalization to multiple soft lines agrees with the results obtained in 
\cite{Creminelli2013a}.

We can check that the formula (\ref{dl-dn-2}) is self-consistent, that is, when we
first let $\ell$ wave numbers go to zero, and next decrease the $\ell+1$ wave number,
we recover the expression (\ref{dl-dn-2}) where we directly take $\ell+1$ soft
wave numbers.
Indeed, the results obtained from the two procedures differ by terms of the form
$(\vk'_{\ell+1}\cdot\vk'_j)/k_j'^2$ that are negligible with respect to the terms
of the form $(\vk_i\cdot\vk'_j)/k_j'^2$.
However, the general expression (\ref{dl-dn-2}) is not a mere consequence of the
iterated use of the equation at $\ell=1$. Indeed, the iterative procedure only applies
when there is a strong hierarchy between the soft wave numbers, 
$k'_1 \ll k'_2 \ll .. \ll k'_{\ell}$, whereas Eq.(\ref{dl-dn-2}) is also valid when the
soft wave numbers are of the same order.

The remarkable property of these relations is that they do not require the 
hard wave numbers $\vk_i$ in Eq.(\ref{dl-dn-2}) to be in the linear or perturbative
regime. In particular, they still apply when these high wave numbers $\vk_i$ are
in the highly nonlinear regime governed by shell-crossing effects and affected
by baryon processes such as star formation and cooling.
The only requirement is the ``scale-separation'' property 
(\ref{dpsi-delta-1})-(\ref{dpsi-k0-1}), which states that long wavelength fluctuations
have a uniform impact on small-scale structures, which are merely transported by
the large-scale velocity flow without deformation, at leading order in the ratio of
scales. We discuss in more details the derivation and the meaning of this property
in Sec.~\ref{examples} below.

In the lowest order case, $\ell=1$ and $n=2$, this gives
\beqa
\lim_{k'\rightarrow 0} B(k',t';k_1,t_1;k_2,t_2)  & = & - P_L(k',t') P(k_1;t_1,t_2) 
\nonumber \\
&& \hspace{-3cm} \times \left( \frac{\vk_1\cdot\vk'}{k'^{2}} 
\frac{\bar{D}_+(t_1)}{\bar{D}_+(t')}  + \frac{\vk_2\cdot\vk'}{k'^{2}} 
\frac{\bar{D}_+(t_2)}{\bar{D}_+(t')} \right) ,
\label{B-1}
\eeqa
where we introduced the bispectrum defined by
\beq
B(k_1,t_1;k_2,t_2;k_3,t_3)= \lag \tdelta(\vk_1,t_1) \tdelta(\vk_2,t_2) 
\tdelta(\vk_3,t_3)\rag' .
\label{B-def}
\eeq

To summarize the derivation above, the consistency relations (\ref{dl-dn-2})
rely on the following conditions:

(a) Gaussian initial conditions, to write Eq.(\ref{tC-tR-1}) ,

(b) the ``scale-separation'' property (\ref{dpsi-k0-1}), to write
Eqs.(\ref{R-1n-2}) or (\ref{R-ln-1}),

(c) the convergence to the linear regime on large scales, to use
Eq.(\ref{dl-dn-2}) rather than Eq.(\ref{dl-dn-1}). This is also a necessary
condition for the property (\ref{dpsi-k0-1}).

\subsection{Multicomponent case}
\label{multi-fluid}

The results obtained in the previous section also apply to cases where there are
several fluids, when their large-scale linear growth rates are identical.
Thus, let us consider $N$ fluids, which may interact with each other and with
gravity (which may be ``modified'' for instance through additional scalar fields
that mediate a fifth force).
Then, each fluid $(\alpha)$ satisfies its own continuity equation,
\beq
\alpha=1, .., N : \;\;\; \frac{\pl \delta^{(\alpha)}}{\pl\tau} 
+ \nabla \cdot [ (1+\delta^{(\alpha)}) \vv^{(\alpha)} ] = 0 .
\label{continuity-N-1}
\eeq
We again assume that decaying or subdominant linear modes have had time to
become negligible with respect to the fastest growing mode, so that we can
define the initial conditions by a single field $\tdelta_{L0}(\vk)$ and in the
linear regime we have
\beq
\tdelta^{(\alpha)}_L(\vk,\tau) = D^{(\alpha)}_+(k,\tau) \, \tdelta_{L0}(\vk) .
\label{deltaL-a-1}
\eeq
(The $k$ dependence arises if we consider modified-gravity scenarios.)
The normalization of $\tdelta_{L0}$ is arbitrary and it is not necessarily equal
to one of the density contrasts or to the total density contrast.
As in Eq.(\ref{v-delta-1}), each linear displacement field obeys
\beq
\tilde{\vPsi}^{(\alpha)}_L(\vk,\tau) = \ii \frac{\vk}{k^2} \, \tdelta^{(\alpha)}_L(\vk,\tau) = 
\ii \frac{\vk}{k^2} \, D^{(\alpha)}_+(k,\tau) \, \tdelta_{L0}(\vk) .
\label{psiL-a-delta0-1}
\eeq
Then, we can follow the derivation presented in Sec.~\ref{cons-rel}.
The only critical point is the assumption (\ref{dpsi-delta-1}), which states that
a large-scale perturbation of $\tdelta_{L0}$ leads to a uniform displacement.
It is clear that this requires the large-scale growing modes $\bar{D}^{(\alpha)}_+(\tau)$
to be identical for all fluids,
\beq
k \rightarrow 0 : \;\;\; D^{(\alpha)}_+(k,\tau) \rightarrow \bar{D}_+(\tau) ,
\label{Dbar-alpha-1}
\eeq
so that a distant large-scale perturbation does not give rise to a local relative velocity
between the different fluids.
[An alternative is for the different fluids to be independent (i.e., they are determined
by the same initial conditions but do not interact), so that we only need each fluid
to respond by its own uniform displacement. 
In the cosmological context, because all
fluids interact through gravity, we only have the possibility (\ref{Dbar-alpha-1}).]
The large-scale common limit (\ref{Dbar-alpha-1}) is satisfied in most cosmological
scenarios, for instance when we consider dark matter and baryons in a
$\Lambda$-CDM universe \cite{Somogyi2010,Bernardeau2013}.
Indeed, on large scales the dominant force is gravity, which acts in the same fashion
on all particle species thanks to the equivalence principle, and we recover the
same linear growing mode that is driven by the gravitational instability.
Effects due to different initial velocities correspond to decaying modes, which
we neglect throughout this paper.
Therefore, in practice the condition (\ref{Dbar-alpha-1}) is not a serious limitation.
Then, Eq.(\ref{dl-dn-1}) becomes
\beqa
\lag \tdelta_{L0}(\vk'_1) .. \tdelta_{L0}(\vk'_{\ell}) \, \tdelta^{(\alpha_1)}(\vk_1,t_1) .. 
\tdelta^{(\alpha_n)}(\vk_n,t_n) 
\rag'_{k_j'\rightarrow 0} = && \nonumber \\
&& \hspace{-6cm} \prod_{j=1}^{\ell}  \left( - P_{L0}(k_j') \sum_{i=1}^n 
\frac{\vk_i\cdot\vk_j'}{k_j'^2} \bar{D}_+(t_i) \right) \nonumber \\
&& \hspace{-6cm} \times \; \lag \tdelta^{(\alpha_1)}(\vk_1,t_1) .. 
\tdelta^{(\alpha_n)}(\vk_n,t_n) \rag' .
\label{dl0-dan-1}
\eeqa
As in Eq.(\ref{dl-dn-2}), this may also be written as
\beqa
\lim_{k_j'\rightarrow 0}  \lag \prod_{j=1}^{\ell} \tdelta^{(\alpha_j')}\!(\vk_j',t_j') 
\prod_{i=1}^n \tdelta^{(\alpha_i)}\!(\vk_i,t_i) \rag' = 
\prod_{j=1}^{\ell} P_L^{(\alpha_j')}\!(k_j',t_j') && \nonumber \\
&& \hspace{-7.9cm} \times \lag \prod_{i=1}^n \tdelta^{(\alpha_i)}(\vk_i,t_i) \rag'  \; 
\prod_{j=1}^{\ell}  \left( \! - \! \sum_{i=1}^n \frac{\vk_i\cdot\vk_j'}{k_j'^{2}} 
\frac{\bar{D}_+(t_i)}{\bar{D}_+(t_j')} \! \right) .
\label{dna-2}
\eeqa
Thus, our approach provides a straightforward generalization to the
multifluid case. In addition to the conditions (a), (b), and (c) given at the end of
Sec.~\ref{cons-rel}, it requires the additional condition 
(\ref{Dbar-alpha-1}):

(d) the linear growing modes of the different fluids have the same large-scale
limit.

The constraint (\ref{Dbar-alpha-1}) agrees with Ref.~\cite{Peloso2013a}, who 
also find that the usual consistency relations no longer hold when there
is a large-scale velocity bias and the linear growth rates of the various fluids
are different. This is also clear from the fact that these consistency relations
express a kinematic effect, that is, how small-scale structures are moved about
by large-scale modes. Then, new terms arise when different fluids respond in
different fashions to large-scale modes \cite{Peloso2013a}.

\subsection{Isocurvature or subdominant modes}
\label{decaying}

In Sec.~\ref{multi-fluid}, as in the single-fluid case described in Sec.~\ref{cons-rel}, 
we assumed
for simplicity that decaying or subdominant linear modes have become negligible,
so that we can focus on the fastest linear growing mode, which also defines
our initial conditions. However, it is also interesting to discuss the case of nonzero
initial isocurvature modes, which correspond to isodensity modes on the
Newtonian scales that we consider.
In standard scenarios, these modes are subdominant with respect to the
adiabatic mode (because the gravitational instability couples all matter
components in the same fashion) and the discussion is similar to the single-fluid
case where we include the decaying mode $\delta_{L-}(\vx,t)$.
This means that in addition to the field $\delta_{L0}(\vx)$, the complete determination
of the initial conditions requires one or several other fields
$\delta_{L0-}^{(i)}(\vx)$. 

For a given value of the decaying or subdominant fields
$\delta_{L0-}^{(i)}$, the analysis of Sec.~\ref{corr-resp} and Eq.(\ref{tC-tR-1})
remain valid, where $\delta_{L0}$ is taken as the dominant linear growing
mode. Then, Eq.(\ref{tC-tR-1}) still holds after we take the average over the
decaying modes $\delta_{L0-}^{(i)}$, provided they are independent from
$\delta_{L0}$. In particular, it is not necessary that these additional fields
be Gaussian. Then, the consistency relations (\ref{dl-dn-1}) and 
(\ref{dl0-dan-1}) remain valid, provided the different fluids have the same
large-scale limit (\ref{Dbar-alpha-1}) for this
dominant linear mode $\delta_{L0}$ and we still have the scale-separation
property (\ref{dpsi-delta-1}) or (\ref{dpsi-k0-1}). 
In the standard cosmological scenario, this remains true for several fluids thanks
to the equivalence principle, which ensures that they respond in the same manner
to the Newtonian gravitational potential.
More precisely, as described in Sec.~\ref{kinematic} below, we can still absorb a
large-scale fluctuation of the linear mode $\delta_{L0}$ through the single
change of coordinate (\ref{xp-def}).

Therefore, the consistency relations in the form (\ref{dl-dn-1}) and (\ref{dl0-dan-1})
remain valid when there are other decaying or subdominant modes (such as
isocurvature modes in multifluid cases).
They would also hold if $\delta_{L0}$ is not the dominant growing mode provided
its large-scale limit (\ref{Dbar-alpha-1}) is again the same for all fluids.
However, in practice we do not measure the linear field $\delta_{L0}$ but only
the matter field $\delta$. Then, the consistency relations in their more
useful form (\ref{dl-dn-2}) and (\ref{dna-2}) only apply in the regime
where $\tdelta(\vk',t') \simeq \bar{D}_+(t') \tdelta_{L0}(\vk')$.
This means that the consistency relations can only be verified by observations
in the regime where decaying and subdominant modes are negligible.

\section{Conditions of validity}
\label{examples}

\subsection{Perturbative check}
\label{perturbative}

The derivation presented in Sec.~\ref{consist} is very general, since it only relies on
Gaussian initial conditions, the linear regime on
large scales, and the scale-separation property (\ref{dpsi-k0-1}).

In particular, it also applies to most modified-gravity scenarios and multifluid systems.
Then, it is interesting to follow in detail how this property appears in an explicit
perturbative treatment of the equations of motion, independently of the form
of the interaction vertices, as long as they respect the conditions above.
For our purpose, we only check the ``squeezed'' bispectrum relation (\ref{B-1})
at the lowest order of perturbation theory.
Following the notations used in Refs.~\cite{Valageas2007,Valageas2008}
for the $\Lambda$-CDM cosmology and Refs.~\cite{BraxPV2012,Brax2013}
for modified-gravity scenarios, we write the equations of motion as
\beq
\cO(x,x') \cdot \tpsi(x') = \sum_{n=2}^{\infty} K_n^s(x;x_1,..,x_n) \cdot 
\tpsi(x_1) .. \tpsi(x_n) ,
\label{O-psi-1}
\eeq
where we introduced the coordinate $x=(\vk,\eta,i)$, where $\eta=\ln[a(t)]$ is the time
coordinate, and $i$ is the discrete index of the $2N$-component vector $\tpsi$.
Here, we consider $N$ fluids, which are described by their continuity equations 
(\ref{continuity-N-1}) and their Euler equations, and focusing on the growing-mode
curl-free velocity component, $\tpsi$ can be written as
\beq
\tpsi(\vk,\eta) = ( \tdelta^{(1)} , -\tilde{\theta}^{(1)}/\dot{a} , \dots , \tdelta^{(N)} , 
-\tilde{\theta}^{(N)}/\dot{a} ) ,
\label{psi-N-def}
\eeq
where $\tilde{\theta}^{(\alpha)}=\nabla\cdot\vv^{(\alpha)}$.
These (matter) fluids are subject to the 
usual Newtonian gravitational potential $\Phi_{\rm N}$ as well as to possible fifth-force
potentials $\Phi^{(\alpha)}$. This includes the case of $f(R)$ theories and 
scalar field models, where using the quasistatic approximation we can write
the additional scalar fields as functionals of the $N$ (matter) density fields
\cite{BraxPV2012,Brax2013}.
Then, if the coupling constants are different or the matter fields interact in a
different manner with the various scalar fields, the new potentials $\Phi^{(\alpha)}$
can be different for the $N$ matter fields.
The linear operator $\cO$ contains the first-order time derivatives $\pl/\pl\eta$
and other linear terms.
The vertices $K_n^s$ are equal-time vertices (within this quasistatic approximation)
of the form
\beqa
K_n^s(x;x_1,..,x_n) & = & \delta_D(\eta_1-\eta) .. \delta_D(\eta_n-\eta) \nonumber \\
&& \hspace{-2.6cm} \times \; \delta_D(\vk_1+..+\vk_n-\vk) \; 
\gamma^s_{i;i_1,..,i_n}(\vk_1,..,\vk_n;\eta) .
\label{Ks-def}
\eeqa
In the standard $\Lambda$-CDM case, the gravitational potential is a linear functional
of the density field, thanks to the Poisson equation, and the nonlinearities only
come from the terms $\nabla\cdot[(1+\delta)\vv]$ and $(\vv\cdot\nabla)\vv$ of the
continuity and Euler equations. Then, the equations of motion are quadratic
and the only nonzero vertices are those given by 
Eqs.(\ref{gam-112})-(\ref{gam-222}) in App.~\ref{perturbative-check}.
In the case of modified-gravity scenarios, or nonlinear fluid interactions, the potentials
$\Phi^{(\alpha)}$ can be nonlinear functionals of the density field that contain terms of
all orders and give rise to vertices $\gamma^s_{2\alpha;2\alpha_1-1,..,2\alpha_n-1}$.
They correspond to source terms, which only depend on the density fields, in 
the Euler equations.

As described in App.~\ref{perturbative-check}, one can solve the equation of motion 
(\ref{O-psi-1}) in a perturbative manner over powers of $\tpsi$.
This allows us to explicitly check, in a very general setting, the bispectrum relation 
(\ref{B-1}) at the lowest order of perturbation theory.
In particular, it shows that this result only relies on two ingredients: 

(a) the linear growth rates of the different fluids coincide in the large-scale limit, 
as in (\ref{Dbar-alpha-1}).

(b) the new vertices $\gamma^{\rm new}$ associated with nonlinear interactions, that
may arise for instance from modified-gravity scenarios (or models of baryonic
physics) must be subdominant with respect to the standard vertices
in the limit $k'\rightarrow 0$ in Eq.(\ref{B0-2}). This means that
$\gamma^{\rm new}_{i;i',i''}(\vk',\vk_2)$ grow more slowly than $1/k'$
for $k'\rightarrow 0$ at fixed $k_2$.

The point (a) was already noticed in Sec.~\ref{multi-fluid} and follows from the
requirement (\ref{dpsi-k0-1}).
The point (b) is satisfied in usual $f(R)$ theories and scalar-field models,
including a nonlinear screening mechanism as for dilaton and symmetron models,
as can be seen from the expressions of the vertices $\gamma^s_{2;1,1}$
given in \cite{Brax2013} (we only need the soft mode $k'$ to be on larger
scales than the range $m^{-1}$ of the scalar field).
This remains valid at the general level, for higher-order vertices and
up to the highly nonlinear regime, and for $n$-point correlation functions,
as discussed in Sec.~\ref{Validity-requirements} below.

\subsection{Validity requirements}
\label{Validity-requirements}

\subsubsection{Separation of scales and kinematic response}
\label{separation}

As described in Sec.~\ref{cons-rel}, in addition to the constraints of Gaussian
initial conditions and recovery of the linear regime on large scales, the
consistency relations only rely on the property
(\ref{dpsi-k0-1}). 
Using Eq.(\ref{delta-q}), the critical property (\ref{dpsi-k0-1}) can also be written
in terms of the nonlinear density contrast as
\beq
k'\rightarrow 0 : \;\;\; \frac{\cD\tdelta(\vk,t)}{\cD\tdelta_{L0}(\vk')} = 
\bar{D}_+(t) \frac{\vk\cdot\vk'}{k'^2} \, \tdelta(\vk,t) .
\label{td1-tdL0-1}
\eeq
Then, we do not need to introduce the displacement field and by substituting 
Eq.(\ref{td1-tdL0-1}) into Eq.(\ref{tR-def}) we directly obtain Eqs.(\ref{R-ln-1})
and (\ref{dl-dn-1}).
This is more general and consistency relations such as Eq.(\ref{dl-dn-1})
hold for any system, beyond the cosmological context, where the
derivative (\ref{td1-tdL0-1}) takes the form of a simple multiplicative factor
in the low-$k$ limit.
An obvious example is the case where the field $\tdelta(\vk)$, which
is no longer interpreted as a density field, is a functional of the form
\beqa
\tdelta(\vk) & = & \exp \biggl [ \sum_{n=1}^{\infty} \int \prod_{i=1}^n \dd\vk_i \;
\delta_D(\vk_1+..+\vk_n - \vk) \nonumber \\
&& \times \, E_n^s(\vk_1,..,\vk_n) \tdelta_{L0}(\vk_1) .. \tdelta_{L0}(\vk_n) \biggl ] ,
\label{example-1}
\eeqa
where the symmetric kernels $E_n^s$ satisfy $E_n^s(0,k_2,..,k_n)=0$ for
$n \geq 2$.

In the cosmological case, the property (\ref{td1-tdL0-1}) means that if we perturb
the initial condition $\tdelta_{L0}$ by a small perturbation
$\Delta\tdelta_{L0}$ that only modifies large-scale linear modes 
[i.e., $\Delta\tdelta_{L0}(\vk')=0$ for $k'>k_c$ where the cutoff $k_c$ is far in the linear
regime and much below the other wave numbers of interest], the nonlinear
density contrast transforms, at linear order over $\Delta\tdelta_{L0}$, as
\beqa
\tdelta_{L0} & \rightarrow & \tdelta_{L0} + \Delta \tdelta_{L0}  \nonumber \\
\tdelta(\vk) & \rightarrow & \tdelta(\vk) + \int \dd\vk' \, \Delta\tdelta_{L0}(\vk') \,
\bar{D}_+(t) \frac{\vk\cdot\vk'}{k'^2} \, \tdelta(\vk) \;\;\;\;\;\; \nonumber \\
&& = \tdelta(\vk) \, e^{\vk\cdot \Delta\vx}  ,
\label{d1-tdL0-1}
\eeqa
with
\beq
\Delta\vx =  \bar{D}_+(t) \int \dd\vk' \, \Delta\tdelta_{L0}(\vk') \,
\frac{\vk'}{k'^2} .
\label{Dx-def}
\eeq
[The last line in Eq.(\ref{d1-tdL0-1}) simply means that $\exp(x)=1+x$ at linear order.]
Then, in configuration space this yields
\beq
\delta(\vx,t) \rightarrow \delta(\vx+\Delta\vx,t) .
\label{d1-tdL0-2}
\eeq
This corresponds to a uniform translation, as was clear from Eq.(\ref{dpsi-k0-1}),
where the displacement field $\vPsi(\vq)$ is modified by a uniform
($\vq$-independent) amount.

Thus, the critical assumption that gives rise to the consistency relations
(\ref{dl-dn-2}) is that, at leading order, a very large-scale perturbation of the initial
conditions only leads to an almost uniform translation of small structures.
This is a hypothesis of scale separation: large scales do not strongly modify  
small-scale structures and only move them around. 
In fact, as noticed above, the hypothesis can be made more general as the leading
order effect does not need to be a uniform shift, it could also be any uniform
multiplicative factor.
If this assumption is satisfied, then the details of the small-scale structures are
not important and the latter can be deep in the nonlinear regime, which is why
the consistency relations (\ref{dl-dn-1}) remain valid when the smaller-scale
wave numbers $k_i$ are in the nonlinear regime.

\subsubsection{Derivation of the kinematic effect}
\label{kinematic}

In the standard cosmological case, the reason why the property (\ref{td1-tdL0-1}),
or equivalently (\ref{dpsi-k0-1}), is valid, is due to the equivalence principle and it 
can be seen as follows, see also \cite{Kehagias2013,Creminelli2013}.
By definition of the functional derivative, an infinitesimal change of the initial condition
$\Delta\delta_{L0}$ leads to a change of the nonlinear displacement field given by
\beq
\Delta \vPsi(\vq) = \int \dd\vk' \, \frac{\cD\vPsi(\vq)}{\cD\tdelta_{L0}(\vk')} \,
\Delta\tdelta_{L0}(\vk') .
\eeq
Therefore, to obtain the low-$k'$ limit of the functional derivative we can
look at a perturbation $\Delta\tdelta_{L0}(\vk')$ that is restricted to $k' < k_c$
with $k_c'\rightarrow 0$. For instance, we can choose a Gaussian perturbation of
size $R \rightarrow \infty$ centered on a point $\vq_c$ at a large distance 
from point $\vq$ ($|\vq_c-\vq| \gg R$).
This limit also means that the distance $|\vq_c-\vq|$ is much greater than the scale 
associated with the transition to the linear regime, so that this localized perturbation 
always remains far away.  
Because we are perturbing the linear growing mode, by definition of the field
$\delta_{L0}$, the perturbation $\Delta\delta_{L0}$ does not correspond to
just adding a mass $\Delta M$ around $\vq_c$. It also means that we are
perturbing the initial velocity field $\vv_{L0}$ by the precise amount that corresponds
to the relation between velocity and density in the growing mode.
In other words, we look at the impact of the change of the linear growing mode
\beq
\delta_L(\vq,\tau) \rightarrow \hat{\delta}_L = \delta_L + \bar{D}_+ \Delta\delta_{L0} , 
\label{hat-deltaL}
\eeq
\beq
\vv_L(\vq,\tau) \rightarrow \hat{\vv}_L =  \vv_L -  \frac{\dd\bar{D}_+}{\dd\tau} 
\nabla_{\vq}^{-1} \cdot \Delta\delta_{L0}
\label{hat-vL}
\eeq
(because $R\rightarrow \infty$ it is the large-scale limit $D_+(k'=0,\tau)$ that appears).
At the linear level, this means that the small-scale region around $\vq$ is falling
towards the distant large-scale mass $\Delta M$ centered on $\vq_c$ as in the
growing-mode regime. In particular, if the fields are everywhere linear, 
we have at once the relation (\ref{psi-psiL-1}), which becomes exact, as well
as the property (\ref{dpsi-k0-1}).
Thus, what we must show is that even when the small-scale region around $\vq$
is nonlinear, the impact of the distant mass $M$ is still to attract the small region
with the same acceleration as in the linear regime, and with negligible tidal effects.
This is most easily seen from the equation of motion of the trajectories $\vx(\vq,\tau)$
of the particles,
\beq
\frac{\pl^2 \vx}{\pl\tau^2} + {\cH} \frac{\pl\vx}{\pl\tau} = - \nabla_{\vx} \Phi
= \vF ,
\label{traj-1}
\eeq
where $\cH=\dd\ln a/\dd\tau$ is the conformal expansion rate and $\Phi$
and $\vF$ are the Newtonian gravitational potential and force.
When we add the perturbation $\Delta M$, the trajectories are modified as
$\vx\rightarrow \hat{\vx}$ and the Newtonian force as 
$\vF  \rightarrow \hat{\vF}$, and they follow Eq.(\ref{traj-1})
with a hat on each field.
In a fashion similar to the method used for inflation consistency relations
\cite{Creminelli2013}, we can
look for a simple solution of this perturbed equation of motion built from the
unperturbed one $\vx(\vq,\tau)$ by a simple transformation.
In our case, we simply need to consider new trajectories $\vx'$ defined by
\beq
\vx'(\vq,\tau) \equiv \vx(\vq,\tau) + \bar{D}_+(\tau) \, \Delta \vPsi_{L0}(\vq) ,
\label{xp-def} 
\eeq
where $\Delta \vPsi_{L0}= -\nabla_{\vq}^{-1} \cdot \Delta\delta_{L0}$
is the perturbation to the linear displacement.
Then, since the unperturbed trajectories obey Eq.(\ref{traj-1}), these auxiliary
trajectories satisfy
\beqa
\frac{\pl^2 \vx'}{\pl\tau^2} + {\cH} \frac{\pl\vx'}{\pl\tau} & = & \vF
+ \left( \frac{\dd^2\bar{D}_+}{\dd\tau^2} + \cH \frac{\dd\bar{D}}{\dd\tau} \right) 
\Delta \vPsi_{L0} \;\;\; 
\label{traj-2-a} \\
& = & \vF'(\vx',\tau) + \Delta \vF_L(\vq,\tau) .
\label{traj-2}
\eeqa
In the second line, we used the relation $\vF'(\vx') = \vF(\vx)$ because the 
uniform translation (\ref{xp-def}) only gives rise to the same translation of the
Newtonian force, since $\vF \propto \nabla^{-1} \cdot \delta$.
The last term follows from Eq.(\ref{traj-1}), which implies at linear order that the
displacement field and the force obey 
\beq
\frac{\pl^2 \vPsi_L}{\pl\tau^2}(\vq,\tau) + {\cH} \frac{\pl\vPsi_L}{\pl\tau}(\vq,\tau) 
= - \nabla_{\vq} \Phi_L(\vq,\tau) = F_L(\vq,\tau) .
\label{traj-3}
\eeq
Then, we note that the auxiliary trajectories $\vx'(\vq,\tau)$ satisfy the same initial
conditions as the perturbed trajectories $\hat{\vx}(\vq,\tau)$, since they coincide
in the linear regime thanks to the construction (\ref{xp-def}).
Moreover, they follow the same equations of motion if we can write 
$\Delta \vF'(\vx',\tau) \simeq \Delta \vF_L(\vq,\tau)$.
This is valid in the limit $R \rightarrow \infty$, because the far-away large-size
region produces a Newtonian force $\Delta\vF$ that varies on scale $R$ and
can be approximated as a constant on the extent of the small-scale region $\vq$
that we consider.
Moreover, since we consider an infinitesimal perturbation $\Delta M$, with
power restricted to wave numbers $k' \rightarrow 0$, the size-$R$ region is
deep in the linear regime and its gravitational potential is set by the Poisson
equation with the linear density $\Delta \delta_L$ as a source term, whence
$\Delta \vF \simeq \Delta \vF_L$.
Therefore, we conclude that $\hat{\vx} = \vx'$ and the effect of the large-size
perturbation $\Delta M$ is to induce the uniform translation (\ref{xp-def}),
which is set by the linear force $\Delta \vF_L$.
This gives the property (\ref{psi-psiL-1}),  and hence the results
(\ref{dpsi-k0-1}) or (\ref{td1-tdL0-1}), which directly lead to the consistency
relations (\ref{dl-dn-1})-(\ref{dl-dn-2}).

To simplify the analysis above, we chose the perturbation $\Delta M$ to be
located at a far-away distance $\vq_c$. Since the result does not depend on
$\vq_c$, this is indeed legitimate, but one may wonder why this is the case.
More precisely, one might think that the result could be different if the perturbation
$\Delta M$ overlaps with the small-scale region $\vq$.
(As in the case of halo bias \cite{Kaiser1984}, 
one could imagine that adding a uniform overdense
background accelerates the collapse and even makes qualitative changes to the
density field.) 
This is not the case, at leading order in the limit $R \rightarrow \infty$, because the
dominant effect is the purely kinematic transformation (\ref{d1-tdL0-2}).
Indeed, the large-scale perturbation gives rise to coupled perturbations 
$\{ \Delta \delta_{L0}, \Delta\vPsi_{L0}, \Delta\vv_{L0}, \Delta\Phi_{L0}, 
\Delta \vF_{L0} \}$, which by definition are related as in the linear growing mode.
Then, from the Poisson equation and the continuity and Euler equations, we
have the scalings $\delta_L \propto \nabla^2 \Phi_L \propto \nabla\cdot\vF_L$
and $\vPsi_L \propto \vv_L \propto \nabla \Phi_L \propto \vF_L$.
Thus, at constant force $\Delta \vF_{L0}$ and velocity $\Delta \vv_{L0}$,
the perturbation to the density scales as
$\Delta\delta_{L0} \sim R^{-1} |\Delta \vF_{L0}|$, which vanishes in the large-scale
limit $R\rightarrow \infty$.
Therefore, in the low-$k'$ limit, a perturbation $\Delta\tdelta_{L0}(\vk')$
corresponds to adding a uniform force field, to which the system reacts by
uniform velocity and displacement fields, while the initial density in the small region
of interest is not perturbed
[and it is merely transported by this uniform flow at $t>0$].
Therefore, at leading order for $k'\rightarrow 0$, we only have the purely kinematic
effect (\ref{d1-tdL0-2}).

In the cosmological context, it is possible to go to the next order over $k'$ beyond
the kinematic consistency relations (\ref{dl-dn-2}), at the price of an 
additional approximation \cite{Valageas2013b,Kehagias2013b}.
To remove the dominant kinematic part that scales as $1/k'$, which is the
focus of this paper, it is convenient to consider spherical averages of the
correlations (\ref{tC-def}). Then, the leading-order terms of the form 
$\vk_i\cdot\vk_j'/k_j'^2$ in Eq.(\ref{dl-dn-2}) vanish as we integrate over the
angles of the vectors $\vk_j'$. Physically, the spherical average means that
the large-scale fluctuation $\Delta\delta_{L0}$ in Eq.(\ref{hat-deltaL}) is
spherically symmetric and does not select any preferred direction.
Then, by symmetry there is no kinematic effect because there is no direction
towards which small scales should be transported.
Therefore, the spherically averaged correlations become sensitive to the 
next-to-leading order effect, of order $k'^0 P_L(k')$ instead of $k'^{-1} P_L(k')$.
This probes the dependence of the small-scale dynamics on a large-scale
uniform density background, or uniform curvature of the gravitational potential.
However, this does not give rise to universal consistency relations such as
(\ref{dl-dn-2}), which derive from the purely kinematic effect (\ref{d1-tdL0-2}),
because the small-scale structures are distorted by a uniform curvature of the 
gravitational potential in a manner that depends on the physical properties of
the system (e.g., the gravitational interaction or the density dependence of
cooling processes if one considers galaxies).
Then, to make some progress one must use approximate symmetries that
relate the dark matter dynamics in different backgrounds but do not apply
to all nonlinear processes, such as galaxy formation 
\cite{Valageas2013b,Kehagias2013b}.

\subsubsection{The equivalence principle as a sufficient condition}
\label{equivalence-principle}

The derivation above might seem a bit superfluous, as the result may look
obvious. However, it helps to explicitly show which ingredients are required
to obtain the consistency relations.
In particular, it is clear that the argument does not depend on the structure
of the small nonlinear object at $\vq$. It can be in the highly nonlinear regime
where complex baryon astrophysical processes (e.g., star formation) are
taking place. Thus, the consistency relations (\ref{dl-dn-1})-(\ref{dl-dn-2})
hold even when the hard wave numbers $k_i$ are in the highly nonlinear regime
and we take into account shell crossing and astrophysical processes 
(star formation, outflows,...).

In the standard case, going back to Newton's equation, 
$m_{\rm I} \ddot{\vx} = - m_{\rm G} \nabla \Phi_{\rm N}$,
where $m_{\rm I}$ and $m_{\rm G}$ are the inertial and gravitational masses  
and $\Phi_{\rm N}$ is Newton's potential, the requirement that the distant
large-scale structure leads to the same displacement for all particles means that
the inertial and gravitational masses are equal.
Thus, in the standard framework, the consistency relations follow from the
equivalence principle, in agreement with the analysis in \cite{Creminelli2013}.
In particular, for the multifluid case discussed in Sec.~\ref{multi-fluid}, 
we explicitly recover the condition (\ref{Dbar-alpha-1}) of identical large-scale
linear growth rates. Indeed, this is the condition to have a unique coordinate
transformation (\ref{xp-def}).

\subsubsection{More general scenarios}
\label{More-general-scenarios}

On a more general setting, to derive the kinematic effect (\ref{d1-tdL0-2})
in Sec.~\ref{kinematic}, we did not explicitly need the Poisson equation and
to specify the potential $\Phi$.
We only needed to recover the linear regime on large scales and to ensure that the force
$\Delta \vF(\vx)$ exerted by a large-scale fluctuation of size $R$ was almost constant
over a smaller-scale region and independent of its small-scale structure.
This means that the consistency relations remain valid
when we include (speculative) long-range forces other than the standard 
Newtonian (more precisely, General Relativity) gravity. 
The only requirement is that a weak form of the equivalence principle
remain valid on large scales.
For instance, we can imagine the following three cases.

(a) There exists a long-range fifth force, $\vF_{\Xi}= -\nabla \cdot \Xi$, which derives
from a potential $\Xi$ that obeys a modified Poisson equation, such as
$(\nabla^2+R_c^2 \nabla^4) \Xi = \delta$. Although this can be seen as a deviation from
General Relativity if we include $\vF_{\Xi}$ in the gravitational interaction,
it obeys the equivalence principle in the sense that we use the same coupling
constant for all matter particles. Thus, we still have Eq.(\ref{traj-1}), with 
$\Phi\rightarrow \Phi+\Xi$, that is,
equality of inertial and gravitational masses, and we recover the
consistency relations as in the standard case because of the equivalence
principle, as in Sec.~\ref{equivalence-principle}.

(b) The equivalence principle can be violated on small scales, associated with
the hard wave numbers $\vk_i$ in Eq.(\ref{dl-dn-2}). 
It is sufficient that the equivalence principle applies in the
large-scale limit, that is, for $k'\rightarrow 0$ or $R\rightarrow \infty$, where $R$
is the size of the distant perturbation $\Delta M$. 
An example would
be modified-gravity scenarios associated with a new scalar field that mediates
a fifth force. At the linear level, this gives rise to modified Newton's constants
$\cG_{\rm N} \rightarrow [1+\epsilon^{(\alpha)}(k,t)] \cG_{\rm N}$ in the equations
of motion of the
matter particles. If different fluids have different couplings to the scalar field,
the factors $\epsilon^{(\alpha)}$ can be different.
However, if they coincide at low $k$ [typical models have $\epsilon(k) \propto k^2$,
which vanishes at $k\rightarrow 0$, as discussed in App.~\ref{modified}], 
the consistency relations remain valid
although the different fluids behave in a different fashion on small scales.
We discuss in more details these scenarios in App.~\ref{modified}.
(In another class of scenarios, such as some coupled dark energy models where
dark matter and baryons show different couplings to the scalar field
\cite{Amendola2004,Saracco2010}, a bias
develops between particle species and the consistency relations are violated
\cite{Peloso2013a}.)

(c) The equivalence principle is violated on all scales, except for the linear
growing mode. Indeed, we only need an almost constant force $\Delta\vF_L$
in Eq.(\ref{traj-2}) (with respect to small-scale structures and particle species)
for the force exerted by a large-scale linear growing-mode 
fluctuation. In principle, we could imagine for instance a scenario where only
the linear growing mode obeys the equivalence principle but arbitrary large-scale
fluctuations do not.
Such an example is given in App.~\ref{toy-model}.
However, this is not expected to be a realistic model and in practice
consistency relations follow from the equivalence principle, as in the
standard case \cite{Creminelli2013}.

\subsection{Galilean invariance}
\label{Gal-Inv-1}

Because the effect of a long-wavelength perturbation is to move the small-scale
structures as in Eq.(\ref{d1-tdL0-2}), the net effect on equal-time density correlations
vanishes, as can be checked in the consistency relation (\ref{dl-dn-1}),
using $\sum_i \vk_i=-\sum_{j}\vk_j'\rightarrow 0$.
The same cancellation for equal-time statistics appears in perturbation theory
computations of the density correlations \cite{Jain1996,Scoccimarro1996}.
This cancels the infrared divergent contributions from different diagrams that
appear if the initial power spectrum has significant power on large scales
(i.e., the variance of the initial velocity is infinite).
In this context, this property is somewhat loosely referred to as ``Galilean invariance'',
by which it is meant that small scales are only transported without deformation
by long-wavelength modes. This terminology refers to the usual case
(in the laboratory or in a static Universe) where the Euler equation reads as
$\pl_t\vv + (\vv\cdot\nabla)\vv= - \nabla \Phi_{\rm N}$, which is invariant through a 
uniform velocity change $\vv \rightarrow \vv+\vv_0$. 
In the case of the expanding universe, using comoving coordinates, the dynamics 
is actually invariant through an extended Galilean transformation (EGT)
\cite{Kehagias2013}, that can be written as
\beq
\vx' = \vx - \vn(\tau) , \;\;\;  \vv' = \vv - \dot{\vn}(\tau) , \;\;\; \delta' = \delta , 
\label{EGT-x-1}
\eeq
\beq
\Phi'_{\rm N} = \Phi_{\rm N} + (\ddot{\vn}+\cH \dot{\vn}) \cdot \vx' ,
\label{EGT-x-2}
\eeq
where the dot denotes the derivative with respect to the conformal time
$\tau=\int \dd t/a$, $\cH = \dot{a}/a$, 
and the shift $\vn(\tau)$ between the primed and unprimed
solutions of the equations of motion is arbitrary.

As pointed out by Ref.\cite{Creminelli2013}, the transformation
(\ref{EGT-x-1})-(\ref{EGT-x-2}) with the specific case $\vn(\tau) = \vn_0 \tau$
is not the reason for the consistency relations (\ref{dl-dn-1})-(\ref{dl-dn-2}),
because it does not have the form of a perturbation to the linear growing mode.
The perturbation that is relevant implies both a change of the velocity field
and of the gravitational potential, with a time-dependent uniform displacement that is
proportional to the linear growing mode $\bar{D}_+(t)$, see 
Eqs.(\ref{hat-deltaL})-(\ref{hat-vL}).
In other words, the consistency relations rely on the invariance of the small-scale
structure (at leading order over $k'$) as it falls towards a distant large-scale
mass $\Delta M$, with its displacement and velocity coupled as in the linear
growing mode, rather than a pure constant velocity boost. 

It is interesting to see through explicit examples that Galilean Invariance
and the validity of the consistency relations are independent properties.

(a) A counterexample, where Galilean Invariance is violated (as well as the
equivalence principle) but the consistency relations are still valid, is provided by
the toy model of App.~\ref{toy-model}.
Through the transformation (\ref{EGT-x-1})-(\ref{EGT-x-2}), we find that the equation
of motion of the fluid component $(\alpha)$ keeps the same form if the gravitational 
potential transforms as
$\Phi'_{\rm N} = \Phi_{\rm N} + \frac{1}{\epsilon^{(\alpha)} }
\left[ \ddot\vn+(\cH+\beta^{(\alpha)})\dot\vn\right] \cdot \vx' .$
This is only possible when the right-hand side does not depend on $(\alpha)$,
that is, when $\vn(\tau) \propto \bar{D}_+(\tau)$ where $\bar{D}_+$ satisfies the
conditions (\ref{D-beta-epsilon}). Thus, in this toy model, the standard Galilean 
Invariance is not satisfied and the Extended Galilean Invariance is satisfied by a single
time-dependent function $\vn(\tau)$
(up to a proportionality factor), which is sufficient to yield the consistency relations.

(b) In the multifluid case, it is possible to build dynamics that obey the
Extended Galilean Invariance (as boosted frames generate new solutions) 
but break the invariance principle and the consistency relations, by choosing
different coupling constants to the gravitational interaction or a fifth-force potential
(e.g., see \cite{Creminelli2013b}).
However, even for a single-component system it is possible to satisfy the
Extended Galilean Invariance while violating the consistency relations (and
the equivalence principle).
Thus, let us consider models with an additional fifth-force long-range potential
$\Xi$ in the modified Euler equation,
$\pl_{\tau} \vv + (\vv\cdot\nabla) \vv + \cH \vv = - \nabla \Phi_{\rm N} - \nabla \Xi$,
and $\Xi[\delta]$ is a functional of the density field.
Then, the Extended Galilean Invariance is satisfied, with the transformations 
(\ref{EGT-x-1})-(\ref{EGT-x-2}) supplemented by $\Xi'=\Xi$.

We may consider two examples,
\beq
({\rm b}1) : \;\;\;   \Xi \propto (\nabla^{-2} \delta)^2 ,  \;\;\; 
\gamma^s_{2;1,1}(\vk_1,\vk_2) \propto \frac{k^2}{k_1^2 k_2^2}  ,
\label{Xi-b1}
\eeq
\beqa
({\rm b}2) : \;\;\;  &&  \Xi \propto (\nabla^{-1} \delta) \cdot  (\nabla^{-1} \delta) , 
\nonumber \\
&& \gamma^s_{2;1,1}(\vk_1,\vk_2) \propto 
\frac{k^2 (\vk_1\cdot\vk_2)}{k_1^2 k_2^2}  ,
\label{Xi-b2}
\eeqa
where $\vk=\vk_1+\vk_2$. In these two cases, the fifth-force potential is quadratic
over the density contrast $\delta$ and this gives rise to a new quadratic vertex
$\gamma^s_{2;1,1}$ in the equation of motion, following the notations of 
Eq.(\ref{Ks-def}). 
Then, going through the check of the bispectrum consistency
relation at the lowest order of perturbation theory, described in 
Sec.~\ref{perturbative}, we find that the new vertex $\gamma^s_{2;1,1}(-\vk',-\vk_2)$
can no longer be neglected as $k'\rightarrow 0$, because it diverges at least
as fast as the $1/k'$ divergence of the standard vertices $\gamma^s_{1;2,1}$
and $\gamma^s_{2;2,2}$.
Therefore, the consistency relations (\ref{dl-dn-1})-(\ref{dl-dn-2}) no longer apply.

We can easily see where the demonstration presented in Sec.~\ref{kinematic}
breaks down.
Let us first consider the model (b1). The force associated with the potential $\Xi$
is $\vF_{\Xi} = - \nabla \Xi \propto \Phi_{\rm N} \; \nabla\Phi_{\rm N}$.
Then, the perturbation to the fifth force due to a distant large-scale perturbation
$\Delta M$ reads at linear order
\beq
\Delta \vF_{\Xi} \propto (\Delta \Phi_{\rm N}) \nabla\Phi_{\rm N} 
+ \Phi_{\rm N} \nabla (\Delta\Phi_{\rm N}) \sim (\Delta\Phi_{\rm N})
\nabla\Phi_{\rm N} ,
\eeq
where we used $\nabla\Phi_{\rm N} \sim \Phi_{\rm N}/r$ and 
$\nabla(\Delta\Phi_{\rm N}) \sim \Delta\Phi_{\rm N}/R$, where $r$ and $R$
are the size of the small object and of the distant large-scale perturbation, with
$r \ll R$.
Thus, the distant large-scale mass $\Delta M$ no longer generates an almost
constant force $\Delta\vF$ over the extent of the small object, because the
slowly varying factor $(\Delta\Phi_{\rm N})$ is modulated by the fast varying
factor $\nabla\Phi_{\rm N}$.
Therefore, we can no longer make the approximation 
$\Delta \vF_L(\vq,\tau) \simeq \Delta\vF'(\vx',\tau)$ in Eq.(\ref{traj-2}) to prove
that the auxiliary trajectories (\ref{xp-def}) are solutions of the perturbed
equations of motion (at lowest order over $k'$). This coupling between
small and large scales is due to the nonlinearity of the potential $\Xi$,
and the same result applies to the model (b2).

This means that both models (b1) and (b2) violate the equivalence principle, in
the sense that two different small-scale structures do not feel the same force
from a distant large-scale fluctuation, which results in the violation of the consistency
relations.
However, there is an additional difference between both models. In the case (b1),
the strong infrared divergence $1/k_i^2$ of the vertex $\gamma^s_{2;1,1}$
actually implies that we do not recover linear theory on large scales.
For instance, the one-loop contribution to the power spectrum arising
from $\lag \psi^{(3)} \psi^{(1)} \rag$, where $\psi^{(n)}$ is the term of order
$n$ of the perturbative expansion over powers of $\delta_L$, scales as
$P_L(k)$ [instead of $k^2 P_L(k)$ in the standard case], because one factor $k^2$,
that arises from the Laplacian of $\Xi$ as we take the divergence of the Euler
equation, is canceled by a denominator $1/k^2$ from a new vertex $\gamma^s_{2;1,1}$.

In the case (b2), the vertex (\ref{Xi-b2}) shows a softer divergence, $\propto 1/k_i$, 
and it actually has the same form as the standard $\gamma^s_{2;2,2}$ of 
Eq.(\ref{gam-222}). Then, linear theory is recovered on large scales, and the
breakdown of the consistency relations is due to the violation of the
equivalence principle.

\section{Conclusion}
\label{Conclusion}

We have presented in this paper a simple nonrelativistic derivation
of the consistency relations that express the $(\ell+n)$ correlation between
$\ell$ soft modes and $n$ hard modes in terms of the correlation
of the hard modes alone, with prefactors that involve the Gaussian power spectrum 
of the soft modes.
This applies to arbitrary numbers of soft wave numbers and fluid components.
This simple derivation explicitly shows that these consistency relations
only rely on three ingredients: (a) Gaussian initial conditions; (b) a scale-separation
property, which states that at leading order large-scale fluctuations merely 
transport small-scale structures without distortions; and (c) the linear regime
is recovered on large scales.

In most of this paper we neglected decaying modes, so that the initial conditions
and large-scale fields are fully specified by a single linear growing mode.
However, we have described in Sec.~\ref{decaying} that the consistency
relations remain valid in the theoretical forms (\ref{dl-dn-1}) and (\ref{dl0-dan-1})
when we include other decaying or subdominant linear modes.
In practice, we do not directly observe each linear mode, which enters these forms
of the consistency relations, but only the total (nonlinear) matter density
contrast. This means that we can only measure these consistency relations
in the regime where the decaying modes are negligible, so that the observed
large-scale density field can be approximated by the linear growing mode.

In agreement with previous works, the critical scale-separation property
that is the basis of the consistency relations follows from the equivalence
principle, as it means that all objects and small-scale structures fall in the
same way in a homogeneous gravitational potential.
In nonstandard scenarios, for instance with a fifth force, the consistency relations
remain valid if (a) the fifth force still obeys the equivalence principle
(e.g., it derives from a potential $\Xi$ that obeys a modified linear Poisson
equation and it shows the same coupling to all particles), or (b) the equivalence
principle is recovered on the scales probed by the soft wave numbers 
[e.g., $k' \ll m$ where $1/m$ is the range of the fifth force mediated by the
additional scalar field, in $f(R)$ or dilaton models].
In a third scenario (c), the equivalence principle can be violated on all scales
except for fluctuations that follow the linear growing mode. However, this is rather
ad-hoc and does not apply to practical cosmological models.

We have also described simple explicit models that obey the Extended Galilean
invariance but violate the consistency relations, because they break the
equivalence principle (through nonlinear effects, which can also preserve or 
prevent the recovery of the linear regime on large scales, depending on the
model).

Because they only involve a kinematic effect, the form of these
consistency relations is very simple and general, and it does not
depend on the details of small-scale physics. They remain valid despite
whatever small-scale nonperturbative processes take place, such as shell crossing
of dark matter trajectories or complex astrophysical processes like star formation
and outflows due to supernovae.
Thus, a detection of a violation of these relations
would signal either non-Gaussian initial conditions, significant decaying mode
contributions, or a modification of gravity that does not converge to General
Relativity on large scales.

These relations become identically zero for equal-time statistics in the
standard scenario (because equal-time statistics cannot distinguish
such uniform displacements). In this perspective, equal-time correlations
could be used to detect deviations from General Relativity if one detects
a nonzero signal \cite{Creminelli2013b}.
If the equivalence principle is satisfied, equal-time statistics are governed
by next order effects, associated with the curvature of the gravitational potential
(as the leading order associated with the constant gradient approximation
vanishes). This distorts the small-scale structures and leads to
more complex and approximate relations that do not share the same level
of generality \cite{Valageas2013b,Kehagias2013b}.

\begin{acknowledgments}

We thank F. Bernardeau, Ph. Brax, and F. Vernizzi for discussions.
This work is supported in part by the French Agence Nationale de la Recherche
under Grant No. ANR-12-BS05-0002.

\end{acknowledgments}

\appendix

\section{Perturbative check}
\label{perturbative-check}

We describe in this appendix the check of the ``squeezed'' bispectrum relation 
(\ref{B-1}) at lowest order of perturbation theory, in a very general
setting that includes a large class of modified-gravity scenarios.
The equation of motion writes as Eq.(\ref{O-psi-1}), with the nonlinear vertices 
(\ref{Ks-def}). 
In the standard $\Lambda$-CDM scenario, the equations of motion are quadratic
and the only nonzero vertices are
\beq
\gamma^s_{2\alpha-1;2\alpha-1,2\alpha}(\vk_1,\vk_2) = 
\frac{(\vk_1+\vk_2)\cdot\vk_2}{2k_2^2} , \label{gam-112}
\eeq
\beq
\gamma^s_{2\alpha-1;2\alpha,2\alpha-1}(\vk_1,\vk_2) = 
\frac{(\vk_1+\vk_2)\cdot\vk_1}{2k_1^2} , 
\label{gam-121}
\eeq
\beq
\gamma^s_{2\alpha;2\alpha,2\alpha}(\vk_1,\vk_2) = 
\frac{|\vk_1+\vk_2|^2 (\vk_1\cdot\vk_2)}{2k_1^2k_2^2} .
\label{gam-222}
\eeq
In the case of modified-gravity scenarios, or nonlinear fluid interactions, the potentials
$\Phi^{(\alpha)}$ can be nonlinear functionals of the density field that contain terms of
all orders and give rise to vertices $\gamma^s_{2\alpha;2\alpha_1-1,..,2\alpha_n-1}$.

Solving the equation of motion (\ref{O-psi-1}) in a perturbative manner, we write
the expansion
\beq
\tpsi = \sum_{n=1}^{\infty} \tpsi^{(n)} , \;\;\; \mbox{with} \;\;\; 
\tpsi^{(n)} \propto (\tdelta_{L0})^n ,
\label{tpsi-n-def}
\eeq
and the first two terms read
\beq
\tpsi^{(1)} = \tpsi_L , \;\;\; \tpsi^{(2)} = R_L \cdot K^s_2 \tpsi_L \tpsi_L ,
\eeq
where $\psi_L$ is the linear growing mode and $R_L$ the linear response function
(i.e., the retarded Green function),
\beq
\cO \cdot \tpsi_L = 0 , \;\;\; \cO \cdot R_L = \delta_D .
\eeq
\beq
\eta_1 < \eta_2 : \;\;\; R_{Li_1,i_2}(k;\eta_1,\eta_2) = 0.
\label{R-causality}
\eeq
The linear growing mode also satisfies
\beq
\eta > \eta' : \;\;\; \tpsi_{Li}(\vk,\eta) = \sum_j R_{Li,j}(k;\eta,\eta') \tpsi_{Lj}(\vk;\eta') ,
\label{psiL-RL}
\eeq
where there is no integration over time.
As in Eq.(\ref{deltaL-a-1}), we also write the linear growing mode as
\beq
\tpsi_i(\vk,\eta) = D_i(k,\eta) \tdelta_{L0}(\vk) , \;\;\; \cO \cdot D = 0
\eeq
where $D_i(k,\eta)$ is the linear growth rate of the $i$-element of the vector
$\tpsi$ and $D=(D_1,..,D_{2N})$. The linear growth rate and the response function
may depend on wave number, depending on the form of the potentials
$\Phi^{(\alpha)}$.

At lowest order, the density bispectrum $B$ reads
\beqa
B(k';k_1\eta_1,k_2,\eta_2) & \equiv & \lag \tdelta_{L0}(\vk') 
\tdelta^{(\alpha_1)}(\vk_1,\eta_1) \tdelta^{(\alpha_2)}(\vk_2,\eta_2) \rag' \nonumber \\
&& \hspace{-2.5cm} = \lag \tdelta_{L0}(\vk') 
\tdelta^{(\alpha_1)(2)}(\vk_1,\eta_1) \tdelta_L^{(\alpha_2)}(\vk_2,\eta_2) \rag' 
+ \mbox{sym.} \nonumber \\
&& \hspace{-2.5cm} = \lag \tdelta_{L0} \; ( R_L \cdot K_2^s \tpsi_L\tpsi_L )_1
\; (\tpsi_L)_2 \rag' + \mbox{sym.} 
\eeqa
where ``sym.'' stands for the symmetric term by $1\leftrightarrow 2$, and we use
simplified notations.
Taking the Gaussian average gives
\beqa
B & = & 2 P_{L0}(k') P_{L0}(k_2) D_{i'}(k',\eta_1') D_{i''}(k_2,\eta_1') 
D_{j_2}(k_2,\eta_2) \nonumber \\
&& \hspace{-0.3cm} \times R_{Lj_1,i}(k_1;\eta_1,\eta_1') 
\gamma^s_{i;i',i''}(-\vk',-\vk_2;\eta_1') + \mbox{sym.} 
\label{B0-2}
\eeqa
where $j=2\alpha-1$ is the component associated with the $(\alpha)$-density.
Next, in the large-scale limit $k'\rightarrow 0$, we are dominated by the vertices
$\gamma^s_{2\alpha-1;2\alpha,2\alpha-1}$ and
$\gamma^s_{2\alpha;2\alpha,2\alpha}$ of 
Eqs.(\ref{gam-121})-(\ref{gam-222}), with
$\gamma^s_{2\alpha-1;2\alpha,2\alpha-1} \simeq 
\gamma^s_{2\alpha;2\alpha,2\alpha} \simeq (\vk_2\cdot\vk')/(2k'^2)$.
[We discuss the nonstandard vertices below Eq.(\ref{B0-6}).]
This yields
\beqa
B_0 & = & \frac{\vk_2\cdot\vk'}{k'^2} P_{L0}(k') P_{L0}(k_2) D_{j_2}(k_2,\eta_2) 
\sum_{\alpha} D_{2\alpha}(k',\eta_1') \nonumber \\
&& \times [ R_{Lj_1,2\alpha-1}(k_1;\eta_1,\eta_1') D_{2\alpha-1}(k_2,\eta_1')
\nonumber \\
&& + R_{Lj_1,2\alpha}(k_1;\eta_1,\eta_1') D_{2\alpha}(k_2,\eta_1') ]
+ \mbox{sym.} 
\label{B0-3}
\eeqa
Using the property (\ref{Dbar-alpha-1}), we can factor the term 
$D_{2\alpha}(k',\eta_1') \rightarrow \bar{D}_2(\eta_1')$ out of the sum.
Here $\bar{D}_2$ is the common large-scale velocity growing mode and
$\bar{D}_1=\bar{D}_+$ is the common large-scale density growing mode.
Then, the sum can be resummed at once from Eq.(\ref{psiL-RL}), using
$k_2\rightarrow k_1$ in the limit $k'\rightarrow 0$.
This gives
\beqa
B_0 \!\! & = & \!\! \frac{\vk_2\cdot\vk'}{k'^2} P_{L0}(k') P_{L0}(k_2) D_{j_2}(k_2,\eta_2)
\bar{D}_2(\eta_1') D_{j_1}(k_1,\eta_1) \nonumber \\
&& + \mbox{sym.} 
\label{B0-4}
\eeqa
Next, we can integrate over the time $\eta_1'$ [because of causality, in the equations
above there was an implicit Heaviside term $\Theta(\eta_1'<\eta_1)$, which arises
from Eq. (\ref{R-causality})], using the continuity equation which implies that
$\bar{D}_2(\eta) = \dd\bar{D}_1(\eta)/\dd\eta$.
This yields
\beqa
B_0 & = & \frac{\vk_2\cdot\vk'}{k'^2} P_{L0}(k') P_{L0}(k_2) 
D_+^{(\alpha_2)}(k_2,\eta_2) \bar{D}_+(\eta_1) \nonumber \\
&& \times D_+^{(\alpha_1)}(k_1,\eta_1) + \mbox{sym.} ,
\label{B0-5}
\eeqa
and using $\vk_2 \rightarrow -\vk_1$,
\beq
B_0 = - P_{L0}(k') P_L^{(\alpha_1,\alpha_2)}(k_1;\eta_1,\eta_2) 
\left[ \frac{\vk_1\!\cdot\!\vk'}{k'^2} \bar{D}_+(\eta_1) \!+\! \mbox{sym.} \right] 
\label{B0-6}
\eeq
This agrees with Eq.(\ref{dl0-dan-1}), and with Eq.(\ref{dna-2}) when we change the 
variable from $\tdelta_{L0}(k')$ to $\tdelta_L(k',\eta')$.
This explicit derivation provides a general check of Eq.(\ref{B-1}) at the lowest
order of perturbation theory.

\section{Modified-gravity scenarios}
\label{modified}

Here we briefly consider the case of modified-gravity models, such as $f(R)$ theories
or scalar field models.
To simplify the analysis we focus on a single matter fluid (we have already seen
the general conditions for multifluid cases above), which feels the usual Newtonian 
gravitational potential $\Phi_{\rm N}$ and an additional fifth-force potential
$\Phi_{\rm A}$. 
For scalar-tensor theories, which involve a new field $\varphi$ that couples to matter
particles through a conformal rescaling of the Jordan-frame metric
\cite{Will:2004nx,Khoury:2003aq,Brax:2008hh,Brax:2012yi,Brax2012b,Brax2013},
$\tilde{g}_{\mu\nu} = A^2(\varphi) g_{\mu\nu}$; this potential reads
$\Phi_{\rm A} = c^2 \ln A(\varphi)$,
while the scalar field obeys the Klein-Gordon equation
$\frac{c^2}{a^2} \nabla^2 \varphi = \frac{\dd V}{\dd \varphi} 
+ \rho \frac{\dd A}{\dd\varphi}$,
where $V(\varphi)$ is the scalar-field potential.
Here we used the quasistatic approximation (as well as the nonrelativistic limit).
In the weak field limit, we can linearize the Klein-Gordon equation around the 
background, $\varphi=\bar\varphi+\delta\varphi$, and we obtain
\beq
\mbox{weak field:} \;\;\; \tPhi_{\rm A} \propto \delta\tilde{\varphi} \propto 
\frac{\delta\trho}{k^2+a^2m^2} ,
\label{dphi-1}
\eeq
where $c^2m^2=\dd^2 V/\dd\bar\varphi^2$ and we consider models where
$A \simeq 1 + \beta\varphi/\MPl$ with $\beta\varphi/\MPl \ll 1$. 
Thus, the total potential $\tPhi=\tPhi_{\rm N}+\tPhi_{\rm A}$ is amplified with
respect to the Newtonian potential by a factor $1+\epsilon$ with
\beq
\epsilon(k,t) \propto \frac{k^2}{k^2+a^2m^2} .
\label{epsilon-def}
\eeq
In very dense objects, a screening mechanism takes place \cite{Khoury:2003aq},
due to the nonlinearities
of the Klein-Gordon equation. As we take $\rho\rightarrow\infty$, at
fixed scale $R$, the left-hand side becomes negligible with respect to each term 
in the right-hand side and the field $\varphi$ in the objects settles down to the solution
of $\dd V/\dd\varphi + \rho \dd A/\dd\varphi=0$ [e.g., for $V=V_0 e^{-\varphi/MPl}$
we have $\varphi \sim \ln (\beta \rho/V_0)$]:
\beq
\mbox{strong field:} \;\;\; \varphi \simeq \varphi_c \;\; \mbox{with} \;\;
\frac{\dd V}{\dd\varphi}(\varphi_c) + \rho \frac{\dd A}{\dd\varphi}(\varphi_c) = 0 .
\label{dphi-2}
\eeq
Then, gradients of the scalar field $\varphi$ and of the potential $\Phi_{\rm A}$ are
negligible and the fifth force vanishes, so that we recover the usual Newtonian
gravity.

As noticed in \cite{Hui:2009kc}, the screening mechanism also means that a
very dense object, which is screened, and a moderate density object, which is 
in the weak-field regime (\ref{dphi-1}), do not feel the same fifth force from a 
given distant object. Indeed, the fifth force due to a distant mass $M$ acts on a
small object at  $\vx$ through the local gradients of the potential $\Phi_{\rm A}$ 
at $\vx$,
and therefore, through the local gradients of the scalar field $\varphi$.
In the weak-field regime (\ref{dphi-1}), the fifth force is proportional to the gravitational
force, with a factor $\epsilon$ that depends on the distance to the mass $M$
($k \sim 1/|\vx'-\vx|$), and does not depend on the small object structure.
This is due to the linear approximation: solutions to the Klein-Gordon equation and
to the potential simply add up.
In contrast, in the strong-field regime (\ref{dphi-2}), the field $\varphi$ is
pinned down to the solution $\varphi_c$, with a very high curvature of the effective
potential $V+\rho A$, and adding a distant mass only gives rise to a small deviation
of the local value of $\varphi$. Then, the fifth force due to the distant object is negligible.
Therefore, moderate-density and high-density objects do not respond in the same
way to the distant mass $M$, which corresponds to a violation of the equivalence
principle \cite{Hui:2009kc}.

Nevertheless, the consistency relation (\ref{dl-dn-2}) remains valid in the
soft mode limit $k' \rightarrow 0$, in the regime $k' \ll a m$. Indeed, 
Eq.(\ref{epsilon-def}) shows that for $k \sim 1/R \rightarrow 0$, in
the weak-field regime for the small object, the fifth force vanishes as $k^2$ 
as compared with the Newtonian gravity. 
This is because Newtonian gravity is a long range force,
with $\tPhi_{\rm N} \sim \tdelta/k^2$, whereas the fifth force is a relatively ``short-range''
force mediated by the scalar field $\varphi$, with a characteristic length
$\sim 1/m$ (realistic models take $1/m \lesssim 1$Mpc$/h$ because of
observational constraints from the Solar System).
This fifth force is also negligible when the small object is in the strong-field regime,
where the screening mechanism makes it insensitive to external fluctuations.
Therefore, the fifth force is subdominant with respect to Newtonian gravity
at leading order in $1/k$ and it does not contribute to the response
(\ref{dpsi-k0-1}) of the small object to a large-scale distant mass, provided
$k' \ll a m$.

Going back to the explicit perturbative check presented in
App.~\ref{perturbative}, this feature explicitly appears as we go from
Eq.(\ref{B0-2}) to Eq.(\ref{B0-3}), where we assume that the new nonlinear
vertices generated by the fifth-force potential are subdominant with respect
to the usual vertices $\gamma^s_{1;2,1}$ and $\gamma^s_{2;2,2}$.
As seen from the explicit expressions given by Eqs.(78)-(79) in 
Ref.\cite{Brax2013}, this is true because the vertices are rational functions
with denominators of the form $1/(k^2+a^2m^2)$ that remain finite
as $k\rightarrow 0$. This is the same denominator as in Eq.(\ref{epsilon-def})
and again it is due to the small-range character of the fifth force.
The same result holds for the $f(R)$ theories, for the same short-range
reason, as can be checked in the explicit expression of the low-order vertices
given by Eqs.(75)-(76) in Ref.\cite{Brax2013}.

\section{Toy model violating the equivalence principle on all scales}
\label{toy-model}

We give here an example of a toy model where the consistency relations
are verified although the equivalence principle is violated.
This relies on the fact that the equivalence principle is recovered for the
specific case of the linear growing mode, which is sufficient to recover the
consistency relations (\ref{dna-2}).
Thus, let us consider the following toy model, made of different particle species
$(\alpha)$ that obey the equations of motion
\beq
\frac{\pl^2\vx^{(\alpha)}}{\pl\tau^2} + \left( \cH + \beta^{(\alpha)}(\tau) \right) 
\frac{\pl\vx^{(\alpha)}}{\pl\tau} = - \epsilon^{(\alpha)}(\tau) \nabla_{\vx} \Phi_{\rm N} ,
\label{toy-1}
\eeq
where 
$\Phi_{\rm N} = 4 \pi \cG_{\rm N} a^2 \nabla^{-2} \sum_{\alpha} \delta\rho^{(\alpha)}$
is Newton's potential.
As compared with the standard case (\ref{traj-1}), we have added a friction term
$\beta^{(\alpha)}$ and an effective Newton's constant $\epsilon^{(\alpha)} \cG_{\rm N}$ 
that depend on the particle species (and on time).
(We could imagine that there is some friction with respect to a noninteracting 
component that exactly follows the Hubble flow and gravity is modified, but this
example is not meant to be realistic.)
This model clearly violates the equivalence principle on all scales when the 
coefficients $\epsilon^{(\alpha)}$ are different.

However, following the procedure described in Sec.~\ref{kinematic}, we can still
build auxiliary trajectories as in Eq.(\ref{xp-def}), with a common displacement
$\bar{D}_+(\tau) \Delta\vPsi_{L0}$ so that all particles move by the same amount
and the potential $\Phi_{\rm N}$ is only displaced without deformation.
Then, the right-hand side of Eq.(\ref{traj-2-a}) contains a term 
$[\dd^2\bar{D}_+/\dd\tau^2+(\cH+\beta^{(\alpha)})\dd\bar{D}_+/\dd\tau]
\Delta\vPsi_{L0}$ that is again identical to $\Delta \vF^{(\alpha)}_L(\vq,\tau)$
if all linear growing modes $\bar{D}_+^{(\alpha)}$ are equal to $\bar{D}_+$.
Using the Poisson and continuity equations and the equation of motion
(\ref{toy-1}) in its linear form, the different linear growing modes are identical
if $\bar{D}_+$ is simultaneously the solution of
\beq
\frac{\dd^2\bar{D}_+}{\dd\tau^2} + \left( \cH+\beta^{(\alpha)} \right)
\frac{\dd\bar{D}_+}{\dd\tau} = \epsilon^{(\alpha)} \frac{3}{2} \cH^2 \Om \bar{D}_+ .
\label{D-beta-epsilon}
\eeq
Choosing for instance for $\bar{D}_+$ the usual solution associated with the coefficients
$\beta^{(\alpha)}=0,\epsilon^{(\alpha)}=1$, we can see that for any set of functions
$\epsilon^{(\alpha)}(\tau)$ we can find functions $\beta^{(\alpha)}(\tau)$ so that
Eq.(\ref{D-beta-epsilon}) is satisfied.
For such a choice, we obtain a toy model that violates the equivalence principle on
all scales, but where the consistency relations (\ref{dl-dn-1})-(\ref{dl-dn-2})
remain valid.
The reason for this is that to derive the consistency relations we only need
the response of small-scale objects to a large-scale perturbation of the linear
growing mode (i.e., the initial conditions). This is not the same thing as adding a
large mass $\Delta M$ far away, because we must modify the density and velocity
fields in a coupled manner.

\bibliography{ref1}   

\end{document}